\documentclass[aps,pra,preprint,showpacs]{revtex4-1}

\usepackage{hyperref, soul}
\usepackage{graphicx}
\usepackage{amsfonts,amsmath,amssymb}
\usepackage{appendix}
\usepackage[latin1]{inputenc}
\usepackage{color}
\usepackage{ulem}
\usepackage{multirow}

\definecolor{forestgreen}{RGB}{34, 139, 34}

\begin{document}

\title{Precision measurement of the branching fractions of the $5p\,\,^2P_{1/2}$ state in $^{88}$Sr$^+$\\ with a single ion in a micro fabricated surface trap}

\author{Jean-Pierre Likforman}

\author{Vincent Tugayé}
\altaffiliation{also at Département de Physique, Ecole Normale Supérieure de Lyon, Université de Lyon, 46 avenue d'Italie, F-69364 Lyon, France}

\author{Samuel Guibal}

\author{ Luca Guidoni}
\email{luca.guidoni@univ-paris-diderot.fr}
\affiliation{Université Paris--Diderot, Sorbonne Paris Cité,\\
Laboratoire Matériaux et Phénomènes Quantiques, UMR 7162 CNRS, F-75205 Paris, France}

\pacs{32.70.Cs, 06.30.Ft, 37.10.Ty}

\begin{abstract} 
We measured the branching fractions for the decay of the $5p\,\,^2P_{1/2}$ state of  $^{88}$Sr$^+$ by applying a recently demonstrated photon-counting sequential method (M.~Ramm {\it et al.}, Phys. Rev. Lett. {\bf 111}, 023004) to a single ion laser-cooled in a micro fabricated surface trap.
The branching fraction for the decay into the $5s\,\,^2S_{1/2}$ ground level was found to be $p=0.9453^{+0.0007}_{-0.0005}$.
This result is in good agreement with recent theoretical calculations but disagrees with previous experimental measurements, however affected by a one order of magnitude larger uncertainty.
This experiment also demonstrates the reliability and the performances of ion micro trap technology in the domain of precision measurements and spectroscopy.

\end{abstract}

\maketitle

\section{Introduction}
\label{sec:intro}

Atomic spectroscopy data are, from an historical point of view, one of the most important experimental inputs that triggered the development of quantum mechanics (e.g. Ängström measurements of Balmer series of the hydrogen atom).
Later on, precision measurements of the characteristic features of atomic transitions (i.e. transition frequencies, levels lifetimes and branching fractions) allowed for the development of theoretical methods that now aim to a complete understanding of atomic level structures, at least in the simpler cases \cite{Safronova:2008}.
The comparison between theory and experiments is then necessary to test these models that are essential for addressing some fundamental questions like parity non conservation or search for electron electric dipole moment \cite{Wood:1997, *Ginges:2004}.   
Precise knowledge of atomic properties is also very important for astronomical and cosmological studies \cite{Bautista:2002} in which easily identified atomic lines give precious information about celestial objects.
Finally, the advent of optical clocks (that display improved performances with respect to atomic microwave clocks that define the time unit) \cite{Margolis:2010} needs precise models in order to obtain reliable evaluations of systematic frequency shifts that affect accuracy (e.g. blackbody radiation shift \cite{Jiang:2009}).
In the case of alkali-earth elements, the singly-ionized state is particularly interesting because theoretical calculations only deal with a single valence electron.
Singly ionized alkali-earth elements are also a system of choice for trapped ion based quantum information experiments \cite{Blatt:2008} and are among the species used for precision clocks \cite{Margolis:2010}.
Therefore, several experimental techniques have been developed that allow for internal and motional quantum state control \cite{Wineland:1998, Blatt:2008, Haffner:2008}.
By restricting ourselves to the case of heavier alkali-earth (i.e. species with $D$ metastable states), these techniques have been recently applied to obtain precision measurements of spectroscopical quantities on laser-cooled $^{40}$Ca$^+$ ions \cite{Barton:2000,Gerritsma:2008, Ramm:2013, Hettrich:2015}, $^{138}$Ba$^+$ ions \cite{Kurz:2008,Auchter:2014, De-Munshi:2015}, and $^{88}$Sr$^+$ ions \cite{Margolis:2004, Barwood:2004, Letchumanan:2005, Lybarger:2011}.
In this paper we present the precision measurement of the branching fractions for the decay of the $5p\,\,^2P_{1/2}$ state of $^{88}$Sr$^+$.
In particular, we measured the probability $p$ and $1-p$ for the decay of the $5p\,\,^2P_{1/2}$ to the $5s\,\,^2S_{1/2}$  and $4d\,\,^2D_{3/2}$ states to be, respectively,   $p=0.9453^{+0.0007}_{-0.0005}$ and $1-p=0.0547^{+0.0005}_{-0.0007}$.
This result can also be expressed in terms of branching ratio $BR=\frac{p}{1-p}$ as $BR=17.27^{+0.23}_{-0.17}$, affected by a fractional uncertainty of $1.3\times 10^{-2}$.
 
Experimental spectroscopy concerning Sr ions has been addressed in several papers, the results of which are compiled in the reference \onlinecite{Sansonetti:2012}.
The experimental transition probabilities $A_{SP}$ and $A_{PD}$ for the $5s\,\,^2S_{1/2}\to 5p\,\,^2P_{1/2}$ ($\nu=711$~THz, $\lambda=422$~nm) and $4d\,\,^2D_{3/2}\to 5p\,\,^2P_{1/2}$ ($\nu=275$~THz, $\lambda=1092$~nm) transitions listed in this compilation (and in the NIST database \cite{Kramida:2015}) are obtained taking into account measurements of the branching fractions and of the lifetime $\tau_{P_{1/2}}$.
Lifetime and branching fractions of $^{88}$Sr$^+$  $5p\,\,^2P_{1/2}$ level have been measured in 1967 by A.~Gallagher in an Argon discharge by Hanle-effect spectroscopy \cite{Gallagher:1967}.
The lifetime of the $5p\,\,^2P_{1/2}$ level has been later measured with increased precision with the fast ion beam technique \cite{Kuske:1978, Pinnington:1995}.
The NIST database is then based on the two measurements: $BR=13.4(2.0)$ \cite{Gallagher:1967} and $\tau_{P_{1/2}}=7.39(7)$~ns \cite{Pinnington:1995}.
A more recent, albeit quite indirect, experimental measurement of $A_{SP}$ is also given in reference \onlinecite{Meir:2014}.

Theoretical works on $^{88}$Sr$^+$ are largely motivated by the use the dipole-forbidden ``clock'' $5s\,\,^2S_{1/2}\to 4d\,\,^2D_{5/2}$ Sr$^+$ transition ($\nu=446$~THz, $\lambda=674$~nm) as a secondary frequency standard \cite{Margolis:2004, Barwood:2004, Madej:2004}.
Indeed, in 2006 the International Committee for Weights and Measures (CIPM) has included this transition among the recommended secondary representation of the second \cite{CIPM:2006}.
The need of exactitude, proper to frequency standards, enforces the accurate calculation of blackbody frequency shift.
In order to calculate such a shift, precise determinations of dipole moments of low lying transition are needed.
Such kind of calculations for  $^{88}$Sr$^+$ have been performed with increasing precision during the last years \cite{Poirier:1993, Biemont:2000, Sahoo:2006, Mitroy:2008, Safronova:2010a}.
Several of these results are resumed and compared in the reference \onlinecite{Jiang:2009}.
Another theoretical calculation of dipole moments can be found in a more recent paper devoted to the estimation of parity non-conservation effects in $^{87}$Sr$^+$ and $^{137}$Ba$^+$ \cite{Dutta:2014}.

The paper is organized as follows.
In section \ref{sec:exp} we present the experimental setup and give some details concerning the surface trap and the implementation of the sequential method.
We present the results in section \ref{sec:res}, we discuss the systematic errors and describe the techniques used to determine their contributions to the final result.
Finally, in section \ref{sec:discuss} we compare the result to the literature and briefly discuss possible improvements.

\section{Experimental methods}
\label{sec:exp}
 \subsection{Trapping, cooling, and laser-locking}

The experiments are based on a symmetric five-wires surface trap \cite{Chiaverini:2005} with a nominal ion-surface distance $d=131$~$\mu$m. 
The trap is micro-fabricated in a cleanroom with standard photolithographic techniques on a silica substrate.
The 5~$\mu$m thick gold electrodes are obtained by electroplating \cite{Allcock:2010} with an inter-electrode distance of $5$~$\mu$m in the central region of the trap.
The chip is glued on a ceramic holder and bonded with 20~$\mu$m diameter gold wires.  
Filtered static voltages (dc) provided by a DAC computer card (Measurement Computing PCI-DAS) feed the ceramic holder through {\it in vacuo}  screened kapton wires.
The trap is driven with a radio-frequency (rf) voltage amplitude $V_{rf}\sim 150$~V at a frequency of 33.2~MHz and typically displays radial frequencies in the 1.5 -- 2~MHz range and an axial frequency of  $\simeq 200$~kHz for Sr$^+$.
The trapping potential is tailored (e.g. tilted) by the application of a set of static voltages to the dc electrodes calculated with the matrix approach developed in reference \onlinecite{Allcock:2010}.
The matrix is derived by the analytical calculation of the electrostatic potential generated by each electrode \cite{House:2008}. 
The stray electric-fields are compensated for using a rf correlation technique \cite{Berkeland:1998} adapted to surface traps \cite{Allcock:2010}.

Sr$^+$ ions are loaded from an oven containing a strontium dendrite (Aldrich, 99.9\% pure).
Neutral atoms are ionized by driving a two-photon transition towards a self-ionizing level \cite{Removille:2009, *Kirilov:2009}.
The photo-ionizing laser pulses are issued from a frequency doubled Ti:Sa oscillator (Tsunami, Spectra-Physics) with a central frequency of 695~THz ($\lambda=431$~nm) and a pulse duration of $\simeq 100$~fs.

\begin{figure}[h]
  \centerline{\includegraphics[width=.25\columnwidth]{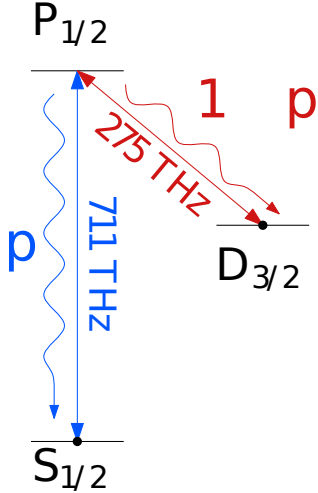}}
  \caption{Low energy levels of $^{88}$Sr$^+$. Two laser sources are used to produce fluorescence cycles of $^{88}$Sr$^+$: a cooling laser at 711~THz (421.7~nm) and a repumping laser at 275~THz (1092~nm). Starting in the $5p\,\,^2P_{1/2}$ energy level, the electronic excitation can either relax to the ground state (with a probability $p$) or to the metastable state $4d\,\,^2D_{3/2}$ (with probability $1-p$).}
  \label{fig:levels} 
\end{figure}

Single trapped $^{88}$Sr$^+$ ions are Doppler cooled using the 711~THz  $5s\,\,^2S_{1/2}\to 5p\,\,^2P_{1/2}$ optical transition (see Fig.~\ref{fig:levels}).
This transition is driven using laser light generated by a commercial extended-cavity GaN laser diode (Toptica DL100). 
The laser frequency is locked to an atomic reference, taking advantage of the near-coincidence ($\nu_{Sr^+}- \nu_{Rb}\simeq  440$~MHz) between the $^{88}$Sr$^+$  $5s\,\,^2S_{1/2}\to 5p\,\,^2P_{1/2}$ and the $^{85}$Rb $5s\,\,^2S_{1/2}(F=2)\to 6p\,\,^2P_{1/2}(F^{\prime}=3)$ transitions \cite{Madej:1998, *Sinclair:2001}.
The 710 962 401 328(40)~kHz absolute frequency of this $^{85}$Rb transition has been recently measured by the frequency-comb technique \cite{Shiner:2007}.
The electronic signal for laser-locking is obtained using a saturated-absorption setup, based on a rubidium cell heated to $100$°~C.
The detuning of the cooling beam with respect to the $5s\,\,^2S_{1/2}\to 5p\,\,^2P_{1/2}$ transition is controlled using an acousto-optic modulator (AOM) in a double-pass geometry driven at a frequency around 220~MHz.
Disregarding the power used for frequency and intensity stabilisation, up to 500~$\mu$W are available at the output of a single-mode polarization-maintaining optical fibre.

A commercial fiber-laser (Koheras Adjustik Y10) drives the $4d\,\,^2D_{3/2}\to 5p\,\,^2P_{1/2}$ 275~THz ``repumping'' transition (see figure~\ref{fig:levels}) to avoid the accumulation of the ions into the metastable $4d\,\,^2D_{3/2}$ state during the cooling process.
This laser has a nominal linewidth of 70~kHz and it is stabilized against long term drifts by a transfer-lock technique using a scanning ring cavity referenced to the stabilized 711~THz laser-diode \cite{Burke:2005, Seymour-Smith:2010}.
The feedback loop of the lock is, in our case, relatively slow (bandwidth $\simeq 3$~Hz) \cite{Dubost:2014}.

A magnetic field of $B$ the order of $1\times 10^{-4}$~T defines a quantization-axis parallel to the substrate making an angle of 45° with the trap axis, orthogonal with respect to the linear polarization of the repumping and cooling laser beams that also propagate along this axis.
This configuration prevents the ions from being optically-pumped into a metastable dark state by the repumping laser alone \cite {Berkeland:2002a}.

Spontaneously emitted 711~THz ("blue") photons are collected by a large-aperture pair of achromatic lenses, spatially filtered with a 150~$\mu$m diameter pinhole, spectrally filtered by an interference filter (Thorlabs FB420-10, 10 nm bandwidth) and detected by a photon-counting photomultiplier head (PMT, Hamamatsu H7828).
The measured global detection efficiency $\epsilon$ of the setup is $\epsilon\simeq1.0\times 10^{-3}$ (see below for the description of the measurement technique).
The logical pulses at the output of the detector are counted and accumulated by a stand alone microcontroller-based gated counter and transferred to the computer that controls the experiment.

Laser beams impinging on the ion are switched-on and -off using AOMs in a double-pass geometry driven through RF switches (Mini-Circuits ZYSWA-2-50DR) and then injected in single mode polarization maintaining optical fibers.
The measured characteristic switching times are in the hundredths of nanoseconds range.
A better than -77~dB extinction ratio has been measured on the repumping beam using a lock-in amplifier: as discussed below such a figure is of importance for the estimate of systematic errors.  
The intensity of the two laser beams impinging on the ion are actively stabilized using the same "noise eater" scheme.
At the output of each fiber the beam passes through a polariser and is then sampled by a beam-splitter and measured by a photodetector.
A gateable analogue PID loop with 10~kHz bandwidth acts on the RF amplitude that drives the AOM in order to keep the measured intensity constant (residual fluctuation  smaller than 5 \%).
Active intensity stabilization allows us to improve the control on resonant Rabi frequencies $\Omega_ 1$ and $\Omega_2$ associated to cooling and repumping  beams respectively.
As explained below, resonant Rabi frequencies, together with respective detunings $\delta_1$ and $\delta_2$, determine the time evolution of the density matrix describing the ion.
In particular, the knowledge of these experimental parameters are needed in order to evaluate systematic errors.
We evaluate the resonant Rabi frequencies of cooling and repumping beams by analyzing a fluorescence spectrum obtained scanning a probe beam at 711~THz in a sequential way similar to that described in reference \onlinecite{Gardner:2014}.
For this analysis, the measurement of the collection efficiency $\epsilon$ helps to reduce uncertainties on the determinations of Rabi frequencies.
The details of this technique, beyond the scope of this paper, will be given elsewhere. 

A fully automated procedure is able to detect an ion loss during the data acquisition: in this case the trap is emptied and a new ion is automatically re-loaded.
This procedure allowed us to compress the effective time needed in order to achieve a low statistical uncertainty.

 \subsection{Sequential acquisition}
We use a sequential technique largely inspired by the one applied for the first time by Ramm and co-workers in order to measure the branching fractions of the $4s\,\,^2S_{1/2}$ state in $^{40}$Ca$^+$ \cite{Ramm:2013}.
The same principle has been used more recently for the measurement of branching fractions of the $6s\,\,^2S_{1/2}$ state in $^{138}$Ba$^+$  \cite{De-Munshi:2015}.
The main differences here, apart from the ion species, are the single-ion operation and the trap technology (micro-fabricated surface trap vs mechanically assembled macro traps). 
Briefly, in our experiment a single $^{88}$Sr$^+$ ion is first Doppler cooled and then prepared in the ground state by switching-off the cooling beam while the repumping beam stays on.
\begin{figure}[h!]
  \centerline{\includegraphics[width=.99\columnwidth]{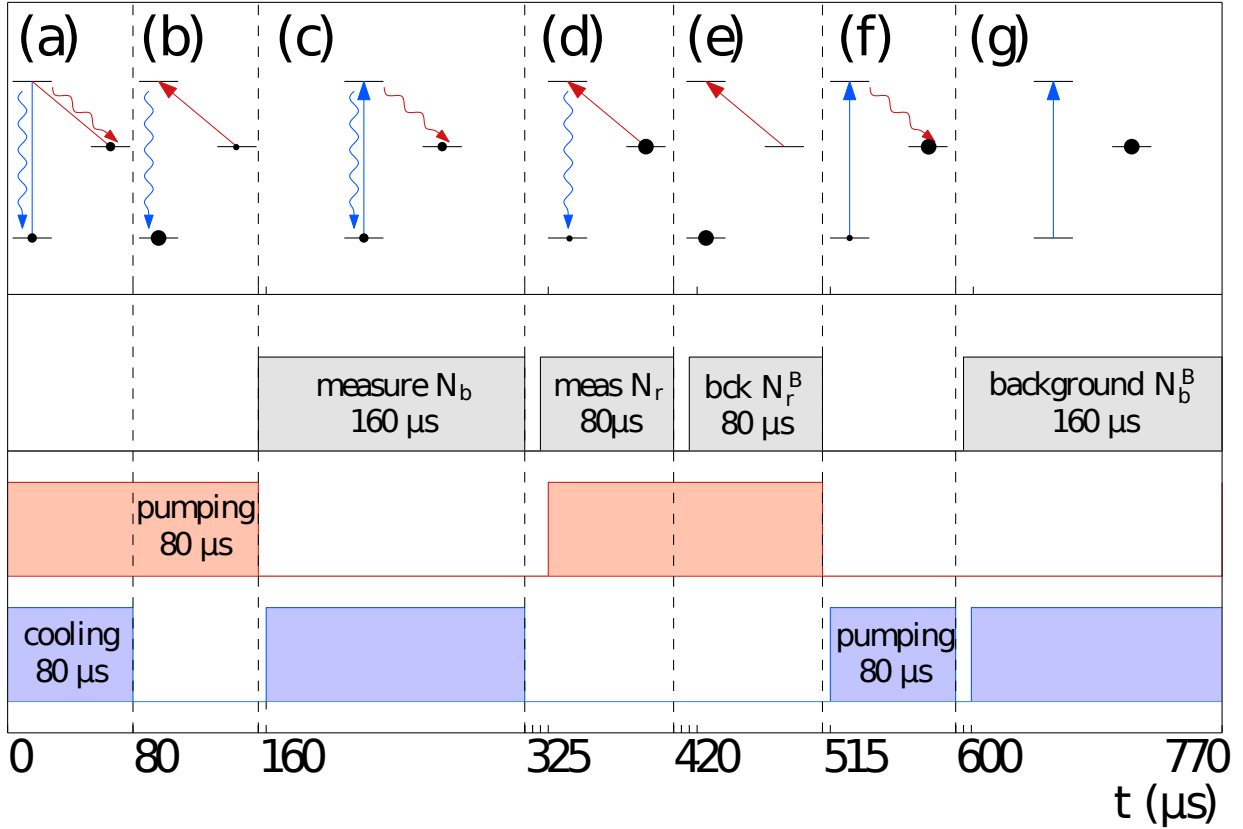}}
\caption{(Color online) Typical acquisition sequence used in a single detection cycle. a) Laser cooling (80~$\mu$s, both lasers are turned on). b) Optical pumping in the ground state (80~$\mu$s, 711~THz laser turned off). c) Measurement of $N_b$ blue photons (160~$\mu$s, 711~THz laser turned on, 275~THz laser turned off). d) Measurement of $N_r$ blue photons (80~$\mu$s, 711~THz laser turned off, 275~THz laser turned on). e) Measurement of the background signal $N_r^B$ associated with phase d) (80~$\mu$s,711~THz laser turned off, 275~THz laser turned on). f) Optical pumping in the metastable $4d\,\,^2D_{3/2}$ state to prepare the measurement of the background signal $N_b^B$ associated with phase c) (80~$\mu$s, 711~THz laser turned on, 275~THz laser turned off). g) Measurement of the background signal $N_b^B$ associated with phase c) (160~$\mu$s,711~THz laser turned on, 275~THz laser turned off). This sequence of duration 770~$\mu$s is typically repeated $\simeq$ 50 millions of times.}
  \label{fig:chrono} 
\end{figure}
A first counting window is then opened during which the cooling beam drives the  $5s\,\,^2S_{1/2}\to 5p\,\,^2P_{1/2}$ transition in the absence of the repumping beam.
In this phase the ion should scatter an average number of $N_{b}$ blue photons ending up in the $4d\,\,^2D_{3/2}$ long-lived metastable state (lifetime $\tau_{D}=435(4)$~ms \cite{Mannervik:1999}).
A second counting window is then opened during which the repumping beam drives the $4d\,\,^2D_{3/2}\to 5p\,\,^2P_{1/2}$ transition in the absence of the cooling beam.
In this phase the ion should scatter a single $N_{r}=1$ blue photon ending up in the ground state, closing in this way a detection loop.
In the absence of photon losses (i.e. for a perfect detection efficiency $\epsilon=1$) the probability $p$ (resp. $1-p$)  for the decay of the $5p\,\,^2P_{1/2}$ to the $5s\,\,^2S_{1/2}$  (resp. $4d\,\,^2D_{3/2}$) state is obtained by measuring  $\frac{N_{b}}{N_{b}+N_{r}}$ (resp. $\frac{N_{r}}{N_{b}+N_{r}}$). 
This relationship still holds in case of imperfect collection efficiency ($\epsilon<1$) because the correction is a common-mode factor for both measurements (e.g. $\frac{\epsilon N_{b}}{\epsilon N_{b}+\epsilon N_{r}}\equiv \frac{N_{b}}{N_{b}+N_{r}}$).
The method is based on the assumption that this behavior is quite robust against variations of experimental conditions (e.g. Rabi frequencies drifts) \cite{Ramm:2013}.
Repeated counts of the number of scattered photons during the counting windows and an independent measurement of the background counts $N^B_{b}$ and $N^B_{r}$ associated to each phase (laser photons scattered by trap surfaces, residual ambient light, photodetector dark counts) allow for the measurement of the branching fractions.
Without considering the systematic effects, the uncertainty is dominated by the statistical error on $N_{r}$.
A typical chronogram used in the experiment is represented in Figure~\ref{fig:chrono}.
In an experiment we acquire many bunches consisting of several hundredths of sequential acquisitions of $N_{b}$ and $N_{r}$ together with the measurements of corresponding backgrounds and we transfer the corresponding sums of detected photons to the computer.

\section{Results}
\label{sec:res}

A typical acquisition run consists of $\geqslant$ 50 millions of sequence cycles that correspond to a "net" acquisition time of the order of 15--20 hours.
We performed two of such runs with different Rabi frequencies and timings in order to check experimentally the estimations of systematics based on the resolution of optical Bloch equations (OBE), as explained below. 
The raw results of the first acquisition (sum of all detected photons in 54'272'970 cycles) are: $N_{b}=1'295'709$, $N_{r}=105'439$, $N^B_{b}=342'349$, $N^B_{r}=50'418$.
For comparison, the results for the second run of 111'200'000 cycles are: $N_{b}=3'521'973$, $N_{r}=244'082$, $N^B_{b}=1'657'903$, $N^B_{r}=136'141$.
Without taking into account the systematic effects (see below) and assuming a Poisson statistics for the photon counting these results give $p=0.9454(6)$ [$BR=17.33(20)$] and $p=0.9453(5)$ [$BR=17.27(17)$], for the first and second run respectively.
The acquisitions also allow for the evaluation of the average detection efficiency of the setup: we obtained $\epsilon=1.01\times10^{-3}$ for the first run and $\epsilon=0.97\times10^{-3}$ for the second run.
 
\subsection{Systematic effects}
Several systematic effects may affect the raw results presented above.
We first consider the residual birefringence of the detection chain that could induce a polarization-sensitive detection efficiency altering the isotropic behavior of the $5s\,\,^2S_{1/2}\to 5p\,\,^2P_{1/2}$ transition \cite{Ramm:2013}.
This effect can be estimated by repeating the experiment with a different orientation of magnetic field in order to evaluate it at the level of the statistical uncertainty of the final result.
We therefore acquired the same amount of data with another orientation of the magnetic field (rotation of 90° in a plane parallel to the trap surface, same magnitude).
The raw results are in this case $N_{b}=1'671'954$, $N_{r}=114'967$, $N^B_{b}=478'305$, $N^B_{r}=46'035$, which give $p=0.9454(5)$, perfectly compatible with the results obtained with the orthogonal orientation.
We can therefore put an upper limit of $5\times 10^{-4}$ to this effect, in terms of uncertainty on $p$.

Collisions and off-resonant excitations of the $4d\,\,^2D_{5/2}$ long-lived metastable state can also be a source of systematic shifts.
In order to evaluate these effects we performed a detailed study of collisions in our experimental setup, following a method similar to that exposed in reference \onlinecite{Barton:2000}.
In particular we recorded the fluorescence of a single ion during a total time of 43 hours with time bins of 5 ms.
A first result of this study is a measurement of the lifetime of the cooled ion in the trap.
Without taking into account extrinsic events (e.g. de-locking of a laser frequency or accidental switching-off), we had to reload a total number of 99 ions, which gives an average lifetime of 1560~s.
\begin{figure}[h]
 \centerline{\includegraphics[width=.99\columnwidth]{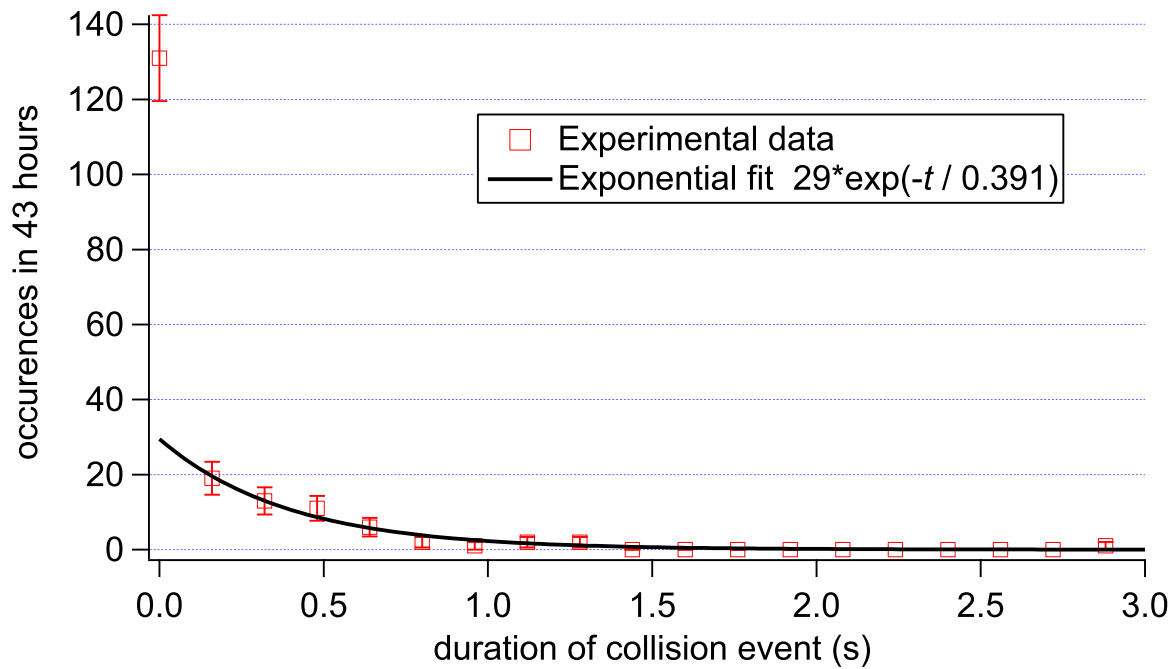}}
 \caption{(Color online) Histogram of time durations of collision events observed during 43 hours of operation in the surface trap. Each event is characterized by an abrupt disappearance of the fluorescence that reappears later in time. The fluorescence is acquired with an integration time of 5~ms and, within this resolution, the reappearance of the fluorescence is also abrupt. The solid line is a fit of the experimental data (with the exclusion of the shortest time-bin of the histogram) with an exponential distribution characterized by an imposed decay time $\tau_{D_{5/2}}=390.8$~ms (fixed parameter) and an amplitude on the first bin (adjustable parameter) of 29 events. The statistical weights of the populations associated to the two classes in this bimodal histogram are 54~\% for the short-lived events and 46~\% for the exponentially-distributed events, respectively.}
\label{fig:histo} 
\end{figure}
The distribution of observed lifetimes is compatible with an exponential distribution.
However, the finite lifetime of the ion in the trap does not affect directly the measurements because we filter out the acquisitions in which the ion is not present.
During the total acquisition time we also observed events displaying an abrupt disappearance of the fluorescence that is later recovered also abruptly.
Following reference \onlinecite{Barton:2000} it is possible to attribute these events to two kinds of phenomena depending on their duration.
In a first case, non-resonant optical pumping and/or fine-structure-changing collisions can bring the ion in the $4d\,\,^2D_{5/2}$ state.
These events should display a duration distributed exponentially with the lifetime $\tau_{D_{5/2}}=390.8$~ms of the $4d\,\,^2D_{5/2}$ state \cite{Letchumanan:2005}.
On the other hand, some of the very short dark periods are likely to be the consequence of smaller perturbations by distant collisions.
This interpretation is supported by the analysis of the histogram of the time durations of the events reported in Fig.~\ref{fig:histo}, in which the two classes of events clearly separate into one fraction following an exponential distribution characterized by $\tau_{D_{5/2}}$ (the lifetime is not an adjustable parameter for the fit displayed with a continuous line) and another fraction, accumulated around the origin, that contributes for 54~\% of the events.
Let us note that the average time that separates these events (1800~s and 1520~s for the long- and short-lived events, respectively) is of the same order of magnitude as what has been observed  by Barton and co-workers under similar pressure conditions \cite{Barton:2000}.
Contrary to the case of reference \onlinecite{Barton:2000}, we do not observe events displaying gradual reappearance of the fluorescence (within our resolution).
A possible explanation of this behaviour resides in the the lower depth of pseudo-potential well in our surface trap.
This characteristic may not allow an ion that reaches high temperatures to stay trapped and being re-cooled with long characteristic times.

The upper limits of the systematic shift induced by these two classes of collisions can be estimated by considering the worst case in which each collision produces a total unbalance in two elementary acquisition cycles (when the ion first "disappears" and then when it "reappears"). 
Because these collision events statistically affect very few acquisition cycles, their role is however marginal: in terms of $p$ the shift is bounded by $2\times 10^{-7}$ (see table~\ref{tab:errorbudget}).

Other systematic effects arise from the dead time of the photomultiplier, the finite lifetime of the $4d\,\,^2D_{3/2}$ state, the finiteness of the durations of acquisition windows, the finiteness of the extinction ratio of AOM switching.
All these effects can be evaluated by solving the OBE that describe the time evolution of the atomic density matrix during an acquisition sequence as a function of driving lasers parameters (i.e. $\Omega_1$, $\Omega_2$, $\delta_1$, $\delta_2$).
We numerically solve the OBE including all the Zeeman sub-levels involved in the experiment (i.e. 8 sub-levels) and taking into account the effect of magnetic field $B$ and laser polarizations.
This multi-level approach allows us to better estimate the characteristic times associated to optical pumping in the experimental conditions. 
We feed the OBE with the raw branching fraction $p=0.9453$ from our experiments as a best first order approximation to calculate systematic shifts

To estimate the systematic shift associated with the dead-time $\tau_{PM}=70$~ns of our photomultiplier, we use the time-dependent solution of the OBE to calculate the conditional probability $q$ that, following a first detection event, another photoelectron is emitted within a 70~ns time window.
Since we know that after the emission of a photon, the ion is in the ground-state, the probability $q$ is given by the following expression:
\begin{equation}
q = 1-\exp\left(-\int_0^{\tau_{PM}} \varepsilon A_{SP} \sigma_{PP}(t) \ {\rm d}t \right), 
\end{equation}
\noindent where $A_{SP}$ is the transition probability for the $5s\,\,^2S_{1/2}\to 5p\,\,^2P_{1/2}$ transition and $\sigma_{PP}(t)$ is the level $5p\,\,^2P_{1/2}$ population at time $t$.
By neglecting losses of more than one photon and the contribution of the background photons, the total number of undetected photons in the measurement of $N_b$ photons is then $qN_b$.
In the nominal experimental conditions $q=1.4\times10^{-4}$.
This underestimation of $N_b$ induces a systematic shift of $(7^{+5}_{-3})\times 10^{-6}$ on $p$ (see table~\ref{tab:errorbudget}).
The uncertainty is evaluated by assuming a (very conservative) relative uncertainty of 20\% on $\tau_{PM}$ and  $\Omega_1$, the two parameters that mostly affects this systematic shift.

\begin{table}
\centering

\begin{tabular}{l c c }
\hline 
\hline
Effect  & \multicolumn{2}{c}{systematic shift on $p$} \\
\hline 
\rule{0pt}{1.5em}
Collisions                    &  $< 2\times10^{-7}$                             &-\\[5pt]

\multirow{2}{*}{PM dead time} &  \multirow{2}{*}{$7 \times 10^{-6}$}   & ${-3 \times 10^{-6}}$ \\
                              &                                          &${+5 \times 10^{-6}}$  \\[5pt]
$D_{3/2}$ lifetime \&         &  \multirow{2}{*}{$1 \times 10^{-5}$}   &${+2 \times 10^{-4}}$  \\
finite windows                &                                          &${-2 \times 10^{-5}}$  \\[5pt]
Laser leaks                   &  $1 \times 10^{-8}$                    &$ \pm 1\times 10^{-8}$ \\[5pt]
\hline 
\multirow{2}{*}{Total }       &  \multirow{2}{*}{$2 \times 10^{-5}$}   &  ${+2 \times 10^{-4}}$\\
                              &                                          &${-2 \times 10^{-5}}$  \\ 
\hline
\hline 
\end{tabular} 

\caption{\label{tab:errorbudget} Systematic errors estimations on the branching fraction $p$ calculated using the solutions of OBE that describe the time evolution of the atomic density matrix during an acquisition sequence as a function of experimentally determined parameters (i.e. driving lasers parameters with nominal values $\Omega_1=2\pi\times8.7$~MHz, $\Omega_2=2\pi\times18$~MHz, $\delta_1=-2\pi\times27.5$~MHz, $\delta_2=2\pi\times80$~MHz, and $B=10^{-4}$~T). The uncertainties on the systematic errors are calculated considering a (conservative) 20\% uncertainty on Rabi frequencies. The nominal value for the dead time of photodetector is used ( $\tau_{PM}=70$~ns) and an uncertainty of 20\% is also supposed in order to evaluate the error bar associated with this contribution.
Calculations based on OBE show that an increase of the duration of the window (f) in Fig.~\ref{fig:chrono} up to 160~$\mu$s would result in a reduced systematic shift and uncertainty on $p$: $(-2^{+18}_{-4})\times10^{-7}$ (contribution of the third row).}
\end{table}

The finite lifetime of the $4d\,\,^2D_{3/2}$ state and the finite duration of the sequence time-windows modify the average number of photons detected in each measurement phase with respect to the ideal case (infinite lifetime, infinite detection an preparation windows).
We can identify two main physical mechanisms responsible for this shift: the state preparation errors and the imperfect shelving in the metastable state that ends up with the ion in the fluorescence cycle during a measurement window.
By comparing the average photon numbers obtained by solving the OBE (that take into account the experimental window durations and the lifetime $\tau_{D}=435$~ms) with the ideal case we obtain an estimate of the systematic shift associated to these effects in our experimental conditions.
The estimated contribution to the systematic shift that affects $p$ is $(1^{+20}_{-2})\times 10^{-5}$ (see table~\ref{tab:errorbudget}).
As in the case of the effect of $\tau_{PM}$, the uncertainty that affects this shift is evaluated by assuming a relative uncertainty of 20\% on $\Omega_1$ which is the parameter that mainly affects the shift.
The analysis of the solutions of the OBE allowed us to find that the relatively large value of this uncertainty is due to the short duration of the pumping phase (f) preceding the measurement phase of $N_b^B$ (g) (see Fig.~\ref{fig:chrono}).
Calculations also show that an increase of the duration of the window (f) up to 160~$\mu$s gives a systematic shift affecting $p$ of $(-2^{+18}_{-4})\times10^{-7}$.

The last class of systematic shifts that we estimate are those induced by the imperfect extinctions of the two lasers.
This shift is obtained by comparing the results of the OBE that describe the experiments with perfect extinction to the case in which the extinction ratio is fixed to  -77~dB as measured in the experiment.
The estimated contribution of the imperfect extinction to the systematic shift that affects $p$ is $(1\pm1) \times 10^{-8}$ (see table~\ref{tab:errorbudget}), dominated by the repumping beam leaks.

It is interesting to note that some effects partially cancel because they affect in a similar (albeit not identical) way signal and background.
As an example, this is the case for the errors due to the relaxation of the shelved electronic excitation during the measurement of $N_{b}$ and $N_b^B$.
The compensation is not perfect because the two measurements do not start with the ion in the same electronic state.
However the transient dynamics in a typical experiment only covers a small fraction of the respective acquisition windows. 

The summary of systematic errors is reported in table~\ref{tab:errorbudget}.
Taking into account these errors the final result for the branching fraction $p$ is $p=0.9453^{+0.0007}_{-0.0005}$ [$BR=17.27^{+0.23}_{-0.17}$].

\section{Discussion}
\label{sec:discuss}

\begin{figure}[h!]
  \centerline{\includegraphics[width=.9\columnwidth]{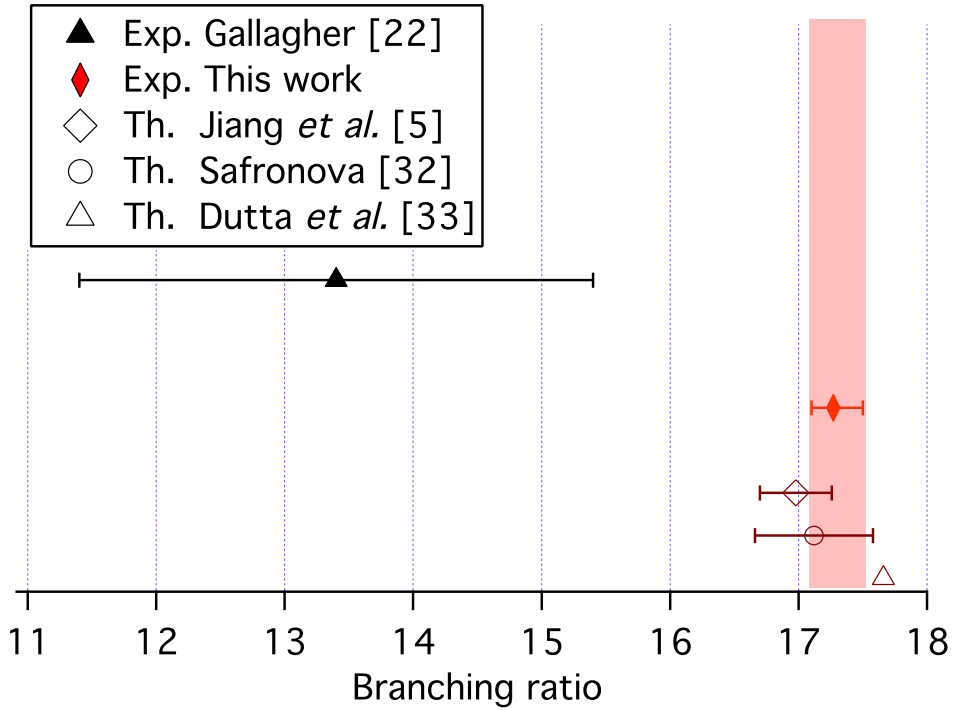}}
  \caption{(Color online) Comparison of our measurement of the branching ratio $BR$ (filled diamond, red) with other experimental measurements 
or theoretical calculations. Vertical axis separation is used to offset the data from different works. The error bars (whenever present) represent the standard error associated to the determination of $BR$. Reference~\onlinecite{Dutta:2014} does not give information about standard error.}
  \label{fig:comp_BR} 
\end{figure}

In this section we discuss how our experimental determination of $BR$ compares to other experimental results and theoretical calculations present in the literature.
First, we compare in Fig.~\ref{fig:comp_BR} the experimental determination of $BR$ obtained by Gallagher \cite{Gallagher:1967} with our result.
As in the case of Ca$^+$ studied in reference~\onlinecite{Gerritsma:2008}, there is no agreement between our data and Gallagher's experiments (performed in an Argon discharge).

We can also compare these results to the theoretical estimates of $BR=A_{SP}/A_{PD}$ that can be obtained starting from the calculated transition probabilities.
The three points on the bottom of Fig.~\ref{fig:comp_BR} (open symbols) have been calculated (with their error bars, whenever applicable) starting from data in references \onlinecite{Jiang:2009,Safronova:2010a,Dutta:2014}.
As outlined in reference \onlinecite{Safronova:2010a}, there is no agreement between recent theoretical calculations and the experimental determination of $BR$ by Gallagher.
This contrasts with the present experimental determination that is indeed compatible, within the smaller error bar, with the calculations of references \onlinecite{Jiang:2009,Safronova:2010a}.

By using the lifetime $\tau_{P}=7.39(7)$~ns measured by Pinnington and co-workers \cite{Pinnington:1995}, the determinations of $p$ can be also recast in terms of transition probabilities $A_{SP}=p/\tau_P$ and $A_{PD}=(1-p)/\tau_P$.
\begin{figure}[h!]
  \centerline{\includegraphics[width=.99\columnwidth]{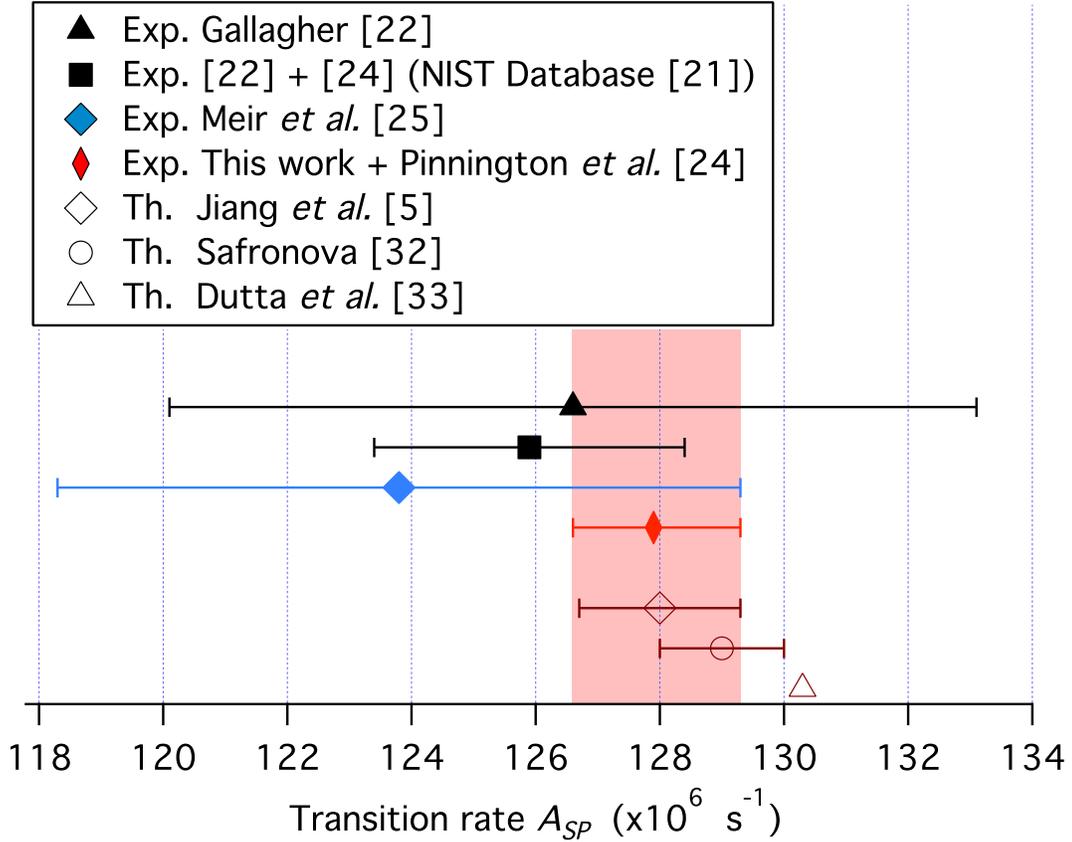}}
  \caption{(Color online) Comparison of measurements and calculations of the transition probability $A_{SP}$. Vertical axis separation is used to offset different measurements.
  In order to }
  \label{fig:ASP} 
\end{figure}
This is the strategy adopted in order to compile the NIST database \cite{Sansonetti:2012,Kramida:2015} that takes advantage of the relatively small uncertainty on $\tau_P$. 
In such a way it is possible to directly test the experimental determinations against the original quantities calculated in theoretical papers. 
In Figure~\ref{fig:ASP} we plot a compilation of the experimental determinations of $A_{SP}$ and the theoretical calculations of the same quantity.
All the determinations are compatible within the uncertainties attributed to measurements or calculations;  let us note that the error bar associated to the present work is dominated by the uncertainty that affects $\tau_P$.
We included in this compilation the results of reference \onlinecite{Meir:2014}, even though the method for the determination of $A_{SP}$  was in this case quite indirect and the exact value of $A_{SP}$ not crucial for their study.
\begin{figure}[h!]
  \centerline{\includegraphics[width=.99\columnwidth]{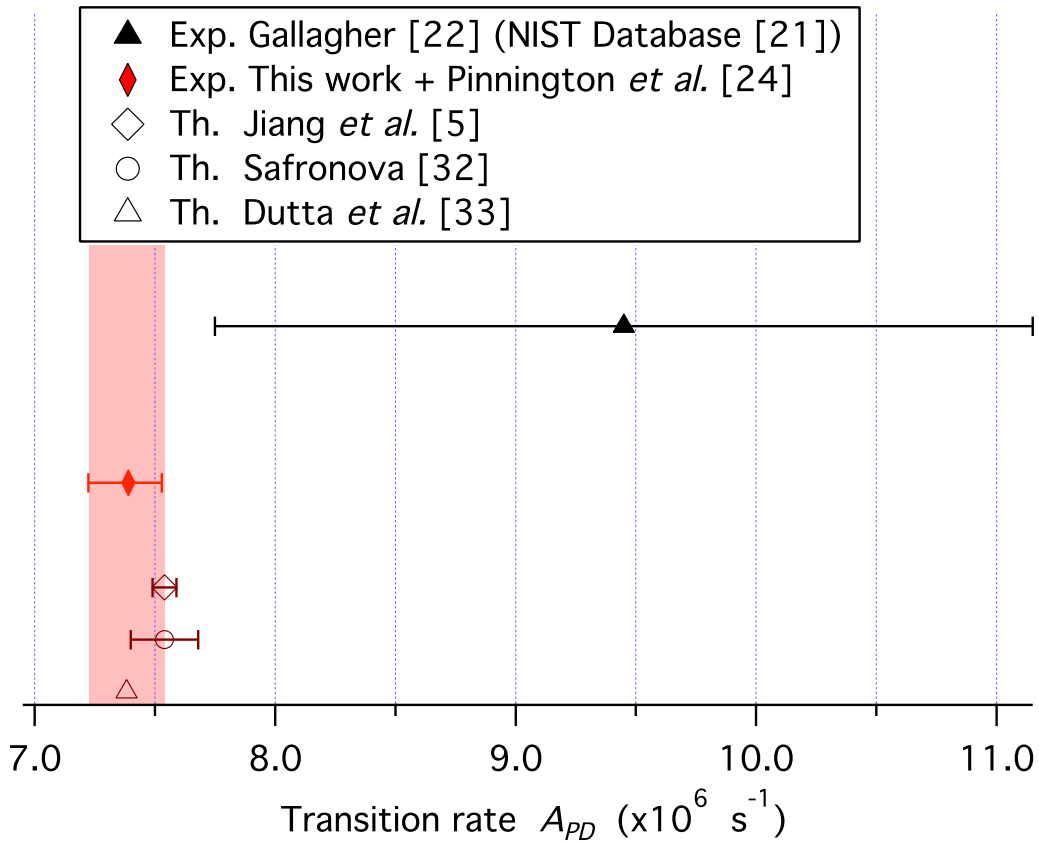}}
  \caption{(Color online) Comparison of measurements and calculations of the transition probability $A_{PD}$.}
  \label{fig:APD} 
\end{figure}

In Figure~\ref{fig:APD} we present the compilation concerning the transition probabilities $A_{PD}$.
The error bar associated to the present work is in this case dominated by the uncertainty that affects $BR$.
It is interesting to note that our work brings back in agreement theory and experimental observations, in a similar way to what is discussed in reference \onlinecite{Hettrich:2015} for $^{40}$Ca$^+$.

It is interesting to analyse the limitations of this method and the possible improvements that could reduce the uncertainty of the present result.
Photon counting statistical uncertainty (dominated by the relatively low total number of photons detected during the $N_r$ measurement phase) gives the largest contribution to our error bar.
Therefore a longer acquisition time will improve the precision of this measurement.
Systematic shift uncertainty could eventually limit this precision.
The design of an optimised time sequence can reduce the uncertainty that has its origin in the finite lifetime of the $4d\,\,^2D_{3/2}$ state and window finiteness down to the  level of $\simeq 1\times10^{-6}$ in terms of fractional uncertainty on $p$.
In this case the main contribution is given by the dead time of the detector.
An improved photon counter (dead time down to $\simeq$~20~ns) and a careful characterization of its dead time could allow for a precision improvement within a factor of ten maintaining realistic acquisition times.
Such a gain, possibly associated with an improved determination of the $5p\,\,^2P_{1/2}$ lifetime, would be interesting in order to put more stringent constraints on theoretical calculations.

In conclusion we measured the branching fractions for the decay of the $5p\,\,^2P_{1/2}$ state of $^{88}$Sr$^+$: the probability $p$ and $1-p$ for the decay of the $5p\,\,^2P_{1/2}$ to the $5s\,\,^2S_{1/2}$  and $4d\,\,^2D_{3/2}$ states are, respectively,   $p=0.9453^{+0.0007}_{-0.0005}$ and $0.0547^{+0.0005}_{-0.0007}$, with a fractional uncertainty (statistical and systematics) on $p$ down to $7\times 10^{-4}$.
In terms of branching ratio the result is $BR=17.27^{+0.23}_{-0.17}$, affected by a fractional uncertainty of $1.3\times 10^{-2}$.
This result can be compared to previous experimental determinations and to theoretical calculations: when considering the branching ratio and the transition probability $A_{PD}$ our work brings back in agreement theory and experimental observations and constitutes an important check for the validity of recent theories.
Finally, this experiment demonstrates the reliability and the performances of ion micro trap technology in the domain of precision measurements and spectroscopy.

\section*{Acknowledgements}
We thank A. Anthore for enlighting discussions concerning trap fabrication; C. Manquest and S. Suffit for their help in cleanroom processes; M. Apfel, P. Lepert and M. Nicolas for technical support.
We also thank B. Szymanski for fruitful discussions and help in the early stages of the experiment.
This study was partly founded by Région Ile-de-France through the DIM Nano-k (projects DEQULOT and EXPLOR@ION).
V. Tugayé thanks the Ecole Normale Supérieure de Lyon for financial support in the form of a fourth year study project.


\end{document}